%% file: paper_la.tex
\documentclass[twocolumn,showpacs,aps,prl,superscriptaddress]{revtex4}

\usepackage{graphicx}
\usepackage{dcolumn}
\usepackage{amsmath}
\usepackage{epsfig}

\input pubboard/babarsym

\input symbols

\newcommand{\BABARPubYear}    {03}
\newcommand{\BABARPubNumber}  {042}

\newcommand{\SLACPubNumber} {10318}

\def\figurebox#1#2#3{%
    \def\arg{#3}%
    \ifx\arg\empty
    {\hfill\vbox{\hsize#2\hrule\hbox to #2{\vrule\hfill\vbox to #1{\hsize#2\vfill}\vrule}\hrule}\hfill}%
    \else
    {\hfill\epsfbox{#3}\hfill}%
    \fi}

\begin{document}


\begin{flushleft}
\babar-PUB-\BABARPubYear/\BABARPubNumber\\
SLAC-PUB-\SLACPubNumber\\
\end{flushleft}

\title{
{\large \bf
Study of \bpsipipm\ and \bpsikpm\ Decays: Measurement
of the Ratio of Branching Fractions and Search for Direct 
\CP\ Violation} 
}

\input pubboard/authors_nov2003.tex

\date{\today}

\begin{abstract}
We study \bpsipipm\ and \bpsikpm\ decays 
in a sample of about $89$ million \BB\ pairs
collected with the \babar\ detector at the \pep2 
asymmetric \B-factory at SLAC. 
We observe a signal of $244 \pm 20$ \bpsipipm\ events and determine the 
ratio \BRpsipipm/\BRpsikpm\ to be 
$[5.37 \pm 0.45 ({\rm stat.}) \pm 0.11 ({\rm syst.})]\%$.
The charge asymmetries for the \bpsipipm\ and 
\bpsikpm\ decays are determined to be
$\calA_{\pi} = 0.123 \pm 0.085 ({\rm stat.}) \pm 0.004 ({\rm syst.})$
and 
$\calA_K = 0.030 \pm 0.015 ({\rm stat.}) \pm 0.006 ({\rm syst.})$, 
respectively.
\end{abstract}


\maketitle
We present an analysis of \bpsipipm\ and \bpsikpm\ decays 
that measures the ratio of branching 
fractions and searches for direct \CP\ violation.
The Cabibbo-suppressed decay \bpsipipm\ proceeds via a $\b\to\c\cbar\d$
transition. It is expected to have a rate about 5\% of that of the 
Cabibbo-allowed mode \bpsikpm. 
The Standard Model predicts that for $\b\to\c\cbar\s$ decays the 
tree and penguin contributions have the same weak phase and thus 
no direct \CP\ violation is expected in \bpsikpm\ decays.
However, for $\b\to\c\cbar\d$, the tree and penguin 
contributions have different phases and charge asymmetries as 
large as a few percent may occur~\cite{Dunietz}.
In the absence of isospin violation, the \CP\ asymmetry 
in \bpsikpm\ provides~\cite{Nir} a measurement of the 
ratio $|\bar{A}/A|$, where $A$ ($\bar{A}$) is the decay 
amplitude for the neutral mode $\Bz (\Bzb) \to \jpsi \KS$. 

Previous studies of the \bpsipipm\ mode have been performed by the
CLEO~\cite{CLEO}, CDF~\cite{CDF}, \babar~\cite{run1anal} 
and Belle~\cite{BELLE} collaborations.
The PDG 2002 average~\cite{PDG2002} of the ratio of branching 
fractions is $(4.2 \pm 0.7) \%$. A recent Belle result 
gives $\BRpsipipm = (3.8 \pm 0.6 \pm 0.3) \times 10^{-5}$. 
The PDG 2002 averages of the charge asymmetries
are $\calA_{\pi} = -0.01 \pm 0.13 $
and $\calA_K = -0.007 \pm 0.019$ (see Eq.~\ref{eq:asym} for 
the definition of the sign of the asymmetry).

The analysis reported in this paper is an update of the \babar\ analysis 
in Ref.~\cite{run1anal} and is based on a larger data set with 
improvements in data reconstruction.
The data were recorded at the \FourS\ resonance with the 
\babar\ detector~\cite{nimpap} at the \pep2 storage ring at the 
Stanford Linear Accelerator Center.
The integrated luminosity is $81.9\invfb$, corresponding to  
$89$ million \BB\ pairs.

At the \babar\ detector, a five-layer silicon vertex tracker (SVT) 
and a 40-layer drift chamber (DCH), in a 1.5-T solenoidal magnetic 
field, provide detection of charged particles and the measurement 
of their momenta.
Electrons are detected in a CsI electromagnetic calorimeter (EMC),
while muons are identified in the magnetic flux return system (IFR), 
which is instrumented with
multiple layers of resistive plate chambers. A ring-imaging Cherenkov
detector (DIRC) with quartz radiators provides charged-particle 
identification. 

We fully reconstruct \bpsihpm\ decays, where $\hpm=\pipm$
or $\Kpm$, from the combination of a \jpsi\ candidate
and a charged track $h^\pm$.
The \jpsi\ candidate is reconstructed via a \psiee\ or \psimm\ decay
and is constrained to the nominal \jpsi\ mass~\cite{PDG2002}.
The electron candidates are combined with reconstructed photons in the 
calorimeter to recover some of the energy lost through bremmstrahlung.
Details of the \jpsi\ reconstruction are given in Ref.~\cite{exclcharm}.
Depending on the final state of the charmonium meson, the \Bpm\ 
candidates are divided into two categories, 
$\B_{ee}$ or $\B_{\mu\mu}$.
The distribution in the angle $\theta_\ell$ in the \jpsi\ rest frame
between one of the daughter leptons $\ell$ of the \jpsi\ and the line
of flight of the recoiling $h^\pm$ is different for signal and background.
The background peaks for $|\cos \theta_\ell |$ near one while the signal
follows a $\sin^{2} \theta_\ell$ distribution.
We require $|\cos \theta_e | < 0.8$ for
$\B_{ee}$ candidates and $|\cos \theta_\mu | < 0.9$ for
$\B_{\mu\mu}$ candidates.

\begin{figure}[!htb]
\begin{center}
  \includegraphics[width=\linewidth]{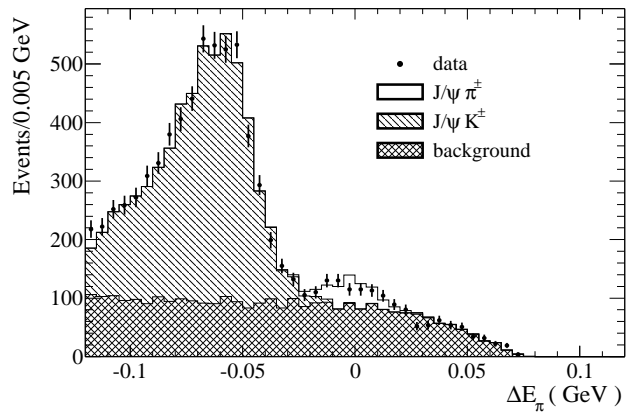}
  \caption{The \deltaepi\ distribution in data (points) compared with 
    the distribution obtained from a simulated experiment (histogram).
    The distributions for each simulated component in the sample,
    normalized to the fitted event yields, are also displayed.
    }
\label{fig:depifit}
\end{center}
\end{figure}

Signal yields and charge asymmetries are determined by
an unbinned maximum likelihood fit to the data. 
A vertex constraint is applied to the reconstructed tracks before
computing the kinematic quantities of the \Bpm\ candidate.
The beam energy-substituted mass \mes\ is defined as
\begin{equation}
\mes =  \sqrt{(s/2+{\bf p}
      \cdot{\bf p_\B})^2 / E^2
  - |{\bf p_\B}|^2} \, ,
\end{equation}
where $\sqrt{s}$ is the total energy of the \epem\ system in the
\FourS\ rest frame, and $(E, {\bf p})$ and $(E_\B, {\bf p_\B})$ 
are the four-momenta of the \epem\ system and the reconstructed 
\B\ candidate, both in the laboratory frame. 
The kinematic variable \deltaepi\ (\deltaek) is defined as 
the difference between the reconstructed energy of 
the \Bpm\ candidate and the beam
energy in the \FourS\ rest frame assuming $\hpm=\pipm(\Kpm)$.
Signal candidates for \bpsipipm\ (\bpsikpm) peak in \mes\ at the \Bpm\ 
meson mass and peak in \deltaepi\ (\deltaek) at $0$.
Candidates are required to satisfy loose requirements on these
variables: $\vert \deltaepi \vert < 120 \mev $, $\vert
\deltaek \vert < 120 \mev  $ and $ \mes > 5.2 \gevcc$. 
The kinematic separation is sufficiently good (see Fig.~\ref{fig:depifit}) 
so that no explicit particle identification is required on the charged hadron 
$h^{\pm}$, thereby simplifying the analysis. 

\begin{figure}[!htb]
\begin{center}
\includegraphics[width=\linewidth]{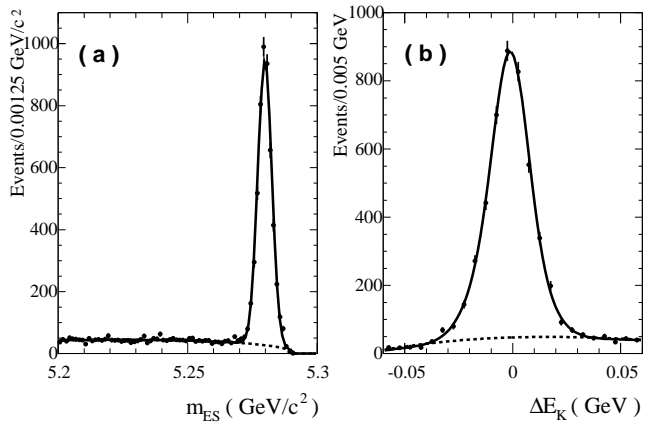}
  \caption{(a) The \mes\ distribution for the \Bpm\ candidates in data. 
  A fit to the sum of a Gaussian and an empirical threshold function
  (dashed curve) is superimposed. The fitted resolution is approximately 
  2.5 \mevcc.
  (b) The \deltaek\ distribution for the \Bpm\ candidates in 
  data with $\mes > 5.27 \gevcc$. A fit to the sum of a
  double Gaussian and a 3rd order polynomial function (dashed curve) is
  superimposed. The fitted resolution is approximately 10.5 \mev}.
\label{fig:resol}
\end{center}
\end{figure}

The selected sample contains $3801$
$\B_{\mu\mu}$ and $4053$ $\B_{ee}$ candidates. 
Figure~\ref{fig:resol}(a) shows the \mes\ distribution in data 
fitted to the sum of a Gaussian and an empirical
phase-space function (Argus function~\cite{ARGUS}) describing the
signal and background components, respectively.
Figure~\ref{fig:resol}(b) shows the \deltaek\ distribution for data
candidates with $\mes > 5.27 \gevcc$ fitted to the sum of a
double Gaussian and a polynomial function, describing the dominant \bpsikpm\
signal and the background contribution, respectively.

The background ($bkg$) from continuum and generic \BB\ decays is characterized 
using events that are outside the signal regions (sidebands of
the data sample). Candidates in the \mes\ sideband are defined by 
the requirement $ 5.20 < \mes < 5.27\gevcc$, where the upper limit 
is approximately four times the experimental resolution below the 
\B\ mass. 
Candidates in the \deltaek\ and \deltaepi\ sidebands are defined by the
requirement $ 42 < \vert \deltaek \vert < 120 \mev$ and 
$ 42 < \vert \deltaepi \vert < 120 \mev$,
where the lower limit is approximately four times the \deltae\ resolution 
obtained from the fit shown in Fig.~\ref{fig:resol}(b). 

We maximize the following extended likelihood function:
\begin{equation*}
L = {\rm e}^{-\sum_{i}N_i}
\prod_{j=1}^M \sum_{i} P_{i}(\alpha^j, \deltaepi^j, p_h^j, \mes^j) \, 
c_i(q^j) \, N_i \,\, ,
\label{eq:likedef}
\end{equation*}
where $j$ is the index of the event, $i$
is the index of the hypothesis ($i=\pi,K,bkg$), $N_i$ is the yield
for each hypothesis, and $M$ is the total number of events in the sample.

The arguments of the probability density functions (PDFs) 
$P_i$ are a discrete variable $\alpha$ that identifies the category
of the \B\ candidate ($\alpha = 1$ for $\B_{ee}$, $\alpha = 2$
for $\B_{\mu\mu}$), and the kinematic observables
$(\deltaepi, p_h, \mes)$, where $p_h$ is the $h^\pm$ momentum in
the laboratory frame.
We assume the same PDFs for \Bp\ and \Bm\ candidates.
If we define $P^{ee}_i(\deltaepi, p_h, \mes)$ and $P^{\mu\mu}_i(\deltaepi, p_h, \mes)$ 
as the PDFs for $\B_{ee}$ and $\B_{\mu\mu}$ candidates, we have
\begin{equation}
P_{i}  = \left\{ \begin{array}{ll}
                   r^{ee}_i P_{i}^{ee} & $\;\;$ \mbox{if $\alpha= 1$}  \\
                   (1-r^{ee}_i) P_{i}^{\mu\mu} & $\;\;$ \mbox{if $\alpha= 2$} \, , 
       \end{array}
	\right .
\end{equation}
where $r^{ee}_i$ is the fraction of $\B_{ee}$ candidates in a given
hypothesis.
In the following we will drop the superscripts $ee$ and $\mu\mu$ when not needed.

The factor $c_i(q)$ is the fraction of candidates with charge $q$ in 
hypothesis $i$:
\begin{equation}
c_i(q) = 
  \left\{ 
        \begin{array}{ll}
        1/2 \, (1-\calA_{i}) & $\;\;$ \mbox{if $q= +1$}\\
        1/2 \, (1+\calA_{i}) & $\;\;$ \mbox{if $q= -1$} \, ,
        \end{array}     
           \right . 
\end{equation} 
where $\calA_{i}$ is the charge asymmetry:
\begin{equation}
\calA_{i} = \frac{N_i^- - N_i^+}{N_i^- + N_i^+} \, . 
\label{eq:asym}
\end{equation} 
The yields $N_i$, asymmetries $A_i$, and fractions $r_i^{ee}$ are 
free parameters in the likelihood fit. 

Since the measured variables $\deltaepi$ and $p_h$ are correlated, 
we define a new set of variables:
\begin{eqnarray*}
D &=& \deltaek - \deltaepi = \gamma \, \left( \sqrt{p_h^2 +
m_K^2}-\sqrt{p_h^2 + m_\pi^2} \, \right) \, , \\
\Sigma &=& (\deltaek + \deltaepi) / (D-a) \, , \\
\Pi &=& D \, ( D/2 - a ) \, ,
\end{eqnarray*}
where $\gamma$ is the Lorentz boost from the laboratory 
frame to the \FourS\ rest frame and $a=240\mev$
is twice the maximum $\vert \deltaepi \vert$ or 
$\vert \deltaek \vert$ value for the data sample.
These variables have the property that $(\deltaepi,D)$ in the pion
hypothesis, $(\deltaek,D)$ in the kaon hypothesis, and $(\Sigma,
\Pi)$ in the background hypothesis are correlated at less 
than the few percent level.
Therefore each
$P_i$ can be written as a product of one-dimensional PDFs:
\begin{eqnarray*}
P_{\pi}(\deltaepi, p_h, \mes ) &=& f_\pi(\deltaepi) g_\pi(D)
h_\pi(\mes) \, , \\
P_{K}(\deltaepi, p_h, \mes ) &=& f_K(\deltaek) g_K(D)
h_K(\mes) \, ,\\
P_{bkg}(\deltaepi, p_h, \mes ) &=& f_{bkg}(\Sigma)
g_{bkg}(\Pi) h_{bkg}(\mes) \, .
\end{eqnarray*}

The $f_\pi$ and $f_K$ components are represented 
by double Gaussians, while $h_\pi$ and $h_K$
are described by single Gaussians.
The parameters of $f_\pi$ and  $h_\pi$ are constrained to be equal to
the parameters of $f_K$ and  $h_K$, respectively.
They are free parameters in the likelihood fit and are extracted 
together with the yields. This strategy reduces
the systematic error due to possible inaccuracies of the 
Monte Carlo (MC) simulation in describing the \deltae\ 
and \mes\ distributions.

The $g_\pi$ and $g_K$ components are each
represented by a phenomenological function with seven fixed
parameters estimated from the MC simulation.
They follow an exponential shape with Gaussian edges.
 
\begin{figure}[!htb]
\begin{center}
  \includegraphics[width=\linewidth]{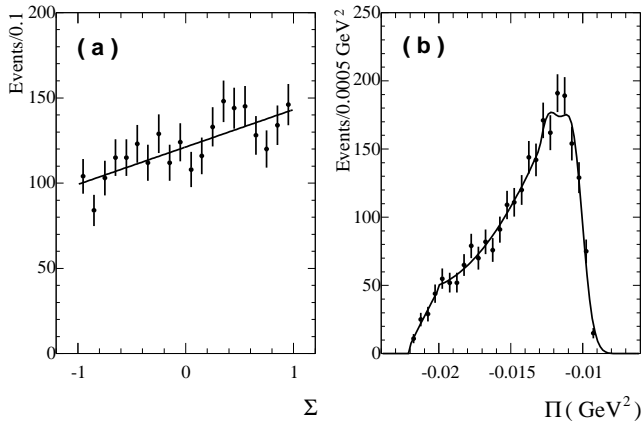}
  \caption{The distribution of (a) $\Sigma$ and (b) $\Pi$ for events 
    in the \mes\ sideband in data. The curve corresponds to the 
projection of the best fit.}
\label{fig:bkgpdf}
\end{center}
\end{figure}

The $f_{bkg}$ component is represented by a linear phenomenological
function with fixed parameters estimated from the
distribution of $\Sigma$ for events in the \mes\ sideband 
(Fig.~\ref{fig:bkgpdf}(a)).

The $g_{bkg}$ component is represented by a phenomenological function 
with twelve fixed parameters, all estimated from the
distribution of $\Pi$ for events in the \mes\ sideband 
(Fig.~\ref{fig:bkgpdf}(b)). 

The $h_{bkg}$ component is represented by the sum of an Argus function
and a Gaussian function, with fixed parameters. The shape parameters are estimated 
from the distribution of \mes\ for events in both the \deltaek\ and \deltaepi\ 
sidebands.
The small number of background events peaking in 
the \mes\ signal region is due to candidates reconstructed from other \bxpsiX\
decays. From detailed MC simulations of inclusive charmonium decays 
we determine $40 \pm 7$ peaking background events in our sample.

The yields determined with the unbinned maximum
likelihood fit to the data sample
are reported in Table~\ref{tab:fitsummary}.
The correlation coefficient between $N_{\pi}$ and $N_{K}$ is $-0.02$.
The probability to obtain a
maximum value of the likelihood smaller than the observed value is
$50 \%$, estimated by MC techniques.
Figure~\ref{fig:depifit} shows the distributions 
of \deltaepi\ for the events in the data,
compared with the distributions obtained by generating events
with a parametric MC simulation based on the PDFs used
in the fit. 

\begin{table}[!htb]
\begin{center}
  \caption{Uncorrected yields $N_i$, fractions of $\B_{ee}$ candidates $r_i^{ee}$ and
    uncorrected charge asymmetries $\calA_{i}$ from the fit to the data sample.
}
\label{tab:fitsummary}
\begin{tabular}{lcccccc}
\hline\hline                
$i$   && $N_i$             && $r_i^{ee} (\%)$      && $\calA_{i}$  \\ \hline
$\pi$ && $\:\:242 \pm 20$  && $50.1 \pm 4.1 $ && $0.117 \pm 0.084$ \\ \hline
$K$   && $4538 \pm 70$     && $46.3 \pm 0.8 $ && $0.028 \pm 0.015$ \\  \hline
$bkg$ && $3074 \pm 60$     && $59.6 \pm 0.9 $ && $0.019 \pm 0.020$ \\
\hline\hline                
\end{tabular}
\end{center}
\end{table}

Possible biases in the likelihood estimates were investigated by
performing the fit on simulated samples of known
composition and of the same size as the data. The samples were 
generated with parametric MC simulations based 
on the PDFs used in the fit. There is no evidence of bias in the
fitted asymmetries, while a less than $1\%$ deviation in the 
fitted yields from the nominal values is present. 
After correcting the yields for the observed bias, we obtain 
$N_{\pi} = 244 \pm 20$, $N_{K} = 4548 \pm 70$, and a ratio
of branching fractions of $(5.37 \pm 0.45) \%$ with an absolute
systematic error of $0.11 \%$.
The dominant source of systematic error is the 
fixed parameters of the PDFs, primarily the PDFs that describe the
background.
Other sources of systematic uncertainty, such as
differences in the reconstruction efficiencies for \psipipm\ and
\psikpm\ events and inaccuracies in the description of the tails
of the \deltae\ resolution function, are found to be negligible.

The sample that is used to determine the charge asymmetries is defined by
imposing as a further requirement that the charged track \hpm\ has a polar 
angle in the range $[0.41,2.54]$ radians, includes
at least $12$ DCH hits, has a momentum in the transverse plane
$p_t > 100 \mevc$, and points back to the nominal interaction point 
within $1.5 \cm$ in the transverse plane and within 
$3 \cm$ along the longitudinal direction. 
For these tracks the difference in tracking efficiency between positively 
and negatively charged tracks - primarily pions - has been studied in
hadronic events by comparing independently the SVT and DCH tracking systems.

The selected sample contains $3902$ \bmpsih\ and $3696$ \bupsih\ 
candidates. From the likelihood fit we obtain the charge asymmetries 
reported in Table~\ref{tab:fitsummary}. The correlation coefficient
between $\calA_{\pi}$ and $\calA_{K}$ is $-0.003$.
Using MC techniques we estimate that the probability to obtain a fitted 
asymmetry $\calA_{K}$ greater or equal to the one observed, 
in the hypothesis of zero asymmetry, is $6.7\%$.

We correct the fitted asymmetries for the small observed difference in 
tracking efficiency between positively and negatively charged tracks,
obtaining $\calA_\pi =  0.123 \pm 0.085 $ and 
$\calA_K   =  0.030 \pm 0.015$.
The uncertainty on the corrections contributes $0.004$
and $0.005$ to the systematic error on $\calA_{\pi}$ and $\calA_K$, 
respectively. 
The asymmetry induced by the different probability
of \Kp\ and \Km\ interactions in the detector material
before the DCH is estimated to be $- 0.004$. This value
is conservatively assumed to be a contribution to the 
systematic uncertainty.
The uncertainty in the fixed parameters of the PDFs, determined 
by fits to simulated or non-signal data sets, contributes 
$0.001$ to the systematic errors on both $\calA_{\pi}$ and
$\calA_K$.

Summing in quadrature statistical and systematic errors, we obtain
a $90\%$ C.L. interval of $[-0.017,0.263]$ for $\calA_{\pi}$ and
$[0.003,0.057]$ for $\calA_K$.

In conclusion we measure the ratio of branching fractions
\begin{equation*}
\frac{\BRpsipipm}{\BRpsikpm} = [5.37 \pm 0.45 ({\rm stat.}) \pm 0.11
({\rm syst.})]\% \, ,
\end{equation*}
which is consistent with theoretical expectations and with previous
measurements.
We also determine the charge asymmetries
\begin{eqnarray*}
\calA_\pi &=&  0.123  \pm 0.085 ({\rm stat.}) \pm 0.004 ({\rm syst.}) \, ,\\
\calA_K   &=&  0.030 \pm 0.015 ({\rm stat.}) \pm 0.006 ({\rm syst.}) \, .
\end{eqnarray*}
Our results are consistent with previous measurements
but with significant improvement in the precision.

\input pubboard/acknow_PRL.tex

\end{document}

%% file: symbols.tex
\def\bpsikpm{\ensuremath{\Bpm \to \jpsi \Kpm}}
\def\bpsipipm{\ensuremath{\Bpm \to \jpsi \pipm}}

\def\BRpsipipm  {\ensuremath{\BR(\bpsipipm)}}
\def\BRpsikpm  {\ensuremath{\BR(\bpsikpm)}}

\def\psiee{\ensuremath{\jpsi \to e^+ e^-}}
\def\psimm{\ensuremath{\jpsi \to \mu^+ \mu^-}}

\def\hpm  {\ensuremath{h^{\pm}}\xspace}

\def\bupsih{\ensuremath{\Bu \to \jpsi h^{+} }}
\def\bmpsih{\ensuremath{\Bub \to \jpsi h^{-} }}
\def\bpsihpm{\ensuremath{\Bpm \to \jpsi h^{\pm} }}

\def\bxpsiX{\ensuremath{B \to \jpsi X}}

\def\psikpm{\ensuremath{\jpsi \Kpm}}
\def\psipipm{\ensuremath{\jpsi \pipm}}

\newcommand{\deltae}{\ensuremath{\Delta E}}
\newcommand{\deltaek}{\ensuremath{\Delta E_K}}

\newcommand{\deltaepi}{\ensuremath{\Delta E_\pi}}

\def\Dp      {\ensuremath{D^+}}
\def\Dm      {\ensuremath{D^-}}
\def\Dstar   {\ensuremath{D^{*}}}
\def\Dstarp  {\ensuremath{D^{*+}}}
\def\Dstarm  {\ensuremath{D^{*-}}}



%% file: pubboard/authors_nov2003.tex
%
\author{B.~Aubert}
\author{R.~Barate}
\author{D.~Boutigny}
\author{F.~Couderc}
\author{J.-M.~Gaillard}
\author{A.~Hicheur}
\author{Y.~Karyotakis}
\author{J.~P.~Lees}
\author{V.~Tisserand}
\author{A.~Zghiche}
\affiliation{Laboratoire de Physique des Particules, F-74941 Annecy-le-Vieux, France }
% ...
\collaboration{The \babar\ Collaboration}
\noaffiliation

%% file: pubboard/acknow_PRL.tex
We are grateful for the excellent luminosity and machine conditions
provided by our \pep2\ colleagues, 
and for the substantial dedicated effort from
the computing organizations that support \babar.
The collaborating institutions wish to thank 
SLAC for its support and kind hospitality. 
This work is supported by
DOE
and NSF (USA),
NSERC (Canada),
IHEP (China),
CEA and
CNRS-IN2P3
(France),
BMBF and DFG
(Germany),
INFN (Italy),
FOM (The Netherlands),
NFR (Norway),
MIST (Russia), and
PPARC (United Kingdom). 
Individuals have received support from the 
A.~P.~Sloan Foundation, 
Research Corporation,
and Alexander von Humboldt Foundation.